\documentclass[aps, prl, twocolumn, superscriptaddress]{revtex4}

\usepackage{graphicx} 
\usepackage{dcolumn} 
\usepackage{bm} 
\usepackage{amsmath}
\usepackage{SIunits}
\usepackage{times}
\usepackage[ansinew]{inputenc}
\usepackage{upgreek}
\usepackage{physics}

\begin{document}

\title{Visualizing dissipative charge carrier dynamics at the nanoscale with superconducting charge qubit microscopy}

\author{Berthold J\"ack}
\affiliation{Princeton University, Joseph Henry Laboratory at the Department of Physics, Princeton, NJ 08544, USA}
\email[electronic address:\ ]{berthold.jack@alumni.epfl.ch}

\date{\today}

\begin{abstract}
The investigation of novel electronic phases in low-dimensional quantum materials demands for the concurrent development of measurement techniques that combine surface sensitivity with high spatial resolution and high measurement accuracy. We propose a new quantum sensing imaging modality based on superconducting charge qubits to study dissipative charge carrier dynamics with nanometer spatial and high temporal resolution. Using analytical and numerical calculations we show that superconducting charge qubit microscopy (SCQM) has the potential to resolve temperature and resistivity changes in a sample as small as $\Delta T\leq0.1\;$mK and $\Delta\rho\leq1\cdot10^{4} \,\Omega\cdot$cm, respectively. Among other applications, SCQM will be especially suited to study the microscopic mechanisms underlying interaction driven quantum phase transitions, to investigate the boundary modes found in novel topological insulators and, in a broader context, to visualize the dissiaptive charge carrier dynamics occurring in mesoscopic and nanoscale devices.
\end{abstract}

\maketitle

\textbf{Introduction.} Over the past decades, a plethora of novel quantum materials has been discovered, in which electronic correlations and a non-trivial bulk topology fundamentally affect their physical properties. Dissipation-less helical edge transport in higher order topological insulators \cite{benalcazar_2017, schindler_2018} and correlated insulating and superconducting states in magic angel twisted bilayer Graphene \cite{cao_2018_1, cao_2018_2} represent only two recent examples, which promise new insights into questions of topological matter and many-body physics, and which carry the prospect of potential technological applications such as dissipation-less electronic charge transport, spintronic devices and topological quantum computation. 

To date, insights on these quantum materials are derived from transport experiments that measure a global resistance drop across a device, or from other experimental techniques, such as photo-electron spectroscopy or scanning tunneling microscopy, which can map out the equilibrium electronic density of states. All these measurements are however inherently insensitive to charge carrier dynamics at small length scales, which dictate the global properties of quantum materials and which will be of relevance for the integration of these material platforms into future electronic devices. The exploration of quantum materials will therefore be accelerated by the concurrent development of new measurement techniques with the ability to measure such local transport properties with high sensitivity and high spatial resolution.
 
Quantum sensing microscopy probes, which harness the sensitivity of a two-level quantum system to perturbing fields from the environment, have started to fill out this gap recently \cite{degen_2017}. Scanning NV center microscopy has proven especially versatile to study spin ordering \cite{gross_2017} and 2D magnetism \cite{thiel_2019}, as well as to measure the temperature and conductivity of metallic surfaces down to temperatures of a few Kelvin with nanometer spatial resolution \cite{kucsko_2013, ariyaratne_2018}. Scanning SQUID on tip thermometry \cite{halbertal_2016}, as a related technique, has mastered the investigation of nano-scale energy dissipation with unprecedented thermal resolution in devices made from 2D materials \cite{uri_2019}. Recent technological advances have also promoted microwave impedance microscopy as a powerful tool \cite{cui_2016} to visualize the topological edge states on the perimeter of an insulating bulk \cite{shi_2019}.

\textbf{Scanning Charge Qubit Microscopy.} Inspired by these advances, we here propose Scanning Charge Qubit Microscopy (SCQM) as a new quantum sensing imaging modality to study dissipative charge carrier dynamics with estimated nanometer spatial and high temporal resolution. SCQM is based on superconducting charge qubits (CQ), such as the Cooper pair box \cite{nakamura_1998}, which are inherently sensitive to charge noise $\delta n(t)$ in the immediate environment. This high sensitivity to charge noise, limiting the CQ's coherence time to about a microsecond \cite{vion_2002, schuster_2005}, renders this type of superconducting qubit less suitable for quantum computation, but on the other hand, makes it a promising candidate for quantum sensing applications, in which charge noise $\delta n(t)$ acting on the CQ can serve as a valuable spectroscopic tool. 

This noise spectroscopy concept we here describe is based on the physical phenomenon that the equilibrium stochastic motion of charge carriers in conducting materials of resistance $R$ gives rise to Johnson-Nyquist voltage noise $\delta V$. At finite temperature $T$ and in the limit of low-frequencies at $k_{\rm B}T\gg\hbar\omega$, this type of noise can be characterized by its voltage noise spectral density $S_{\rm{V}} =4k_{\rm{B}}TR$ ($k_{\rm{B}}$ - Boltzman's constant, $\hbar$ - Planck's constant) \cite{johnson_1928, nyquist_1928}. Measuring Johnson noise in or out of equilibrium, e.g. in the absence or presence of an external drive current $I$ giving rise to the dissipative motion of charge carriers, allows to characterize a sample's resistance, and to quantify the underlying charge carrier dynamics by tracking temperature changes through scattering-induced energy dissipation. In addition, such measurements can also help to distinguish between different transport regimes, such as diffusive and ballistic transport \cite{kolkowitz_2015}. 

\begin{figure*}
\centering
\includegraphics[height=5.5cm, width=17.2cm]{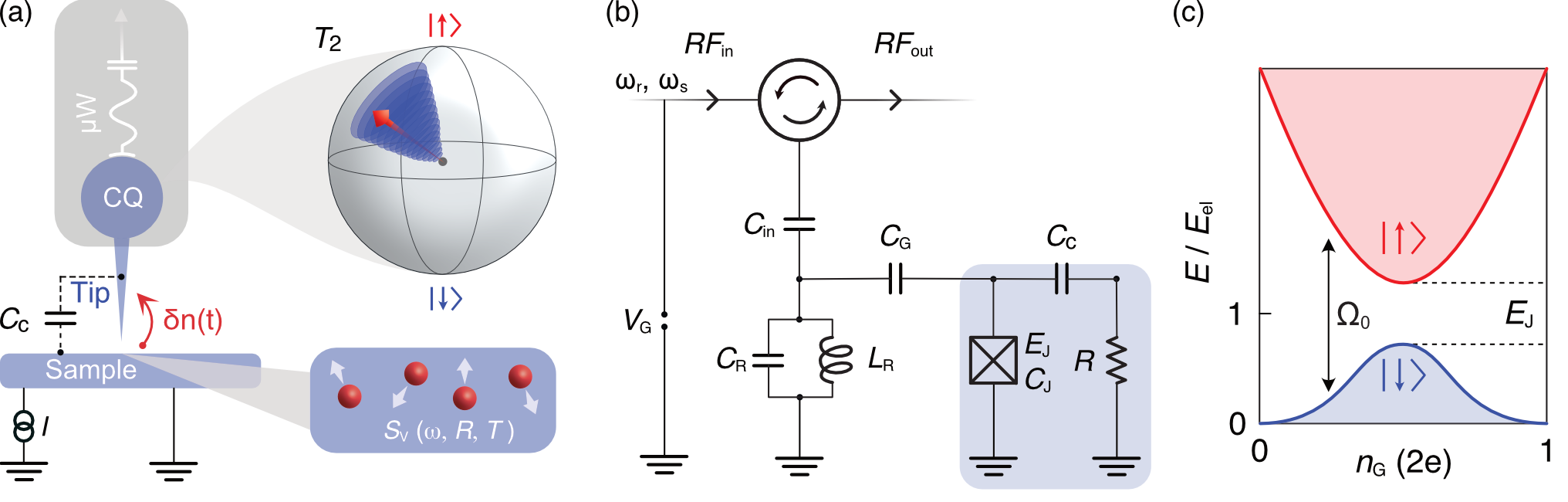}
\caption{\textbf{SCQM - Sensing decoherence with superconducting charge qubits.} (a) Concept sketch of SQCM that uses a superconducting charge qubit (CQ) to sense voltage fluctuations described by a power spectral density $S_{\rm{V}} (\omega, R, T)$ in a sample of interest. The voltage fluctuations are converted into charge fluctuations $\delta n(t)$ by means of a coupling capacitance $C_{\rm{C}}$. Charge fluctuations induce decoherence of the qubit state, which is characterized by a decoherence time $T_2$ and can be read out via microwave ($\mu$W) techniques. Local probe capabilities can be realized using a tip, which is galvanically coupled to the qubit and defines a local geometric coupling capacitance $C_{\rm{C}}$ to the sample. The setup allows for (non-)equilibrium measurements of dissipative transport characteristics using either a grounded sample or by applying a current bias $I$ to the sample. (b) Electrical circuit diagram of SCQM. SQCM is centered around a Josephson junction ($E_{\rm{J}}, C_{\rm{J}}$), which is capacitively coupled ($C_{\rm{G}}$) to a coplanar microwave strip-line resonator (CPS) characterized by its capacitance $C_{\rm{r}}$ and inductance $L_{\rm{r}}$. The gate charge $n_{\rm{G}}$ can be adjusted through the gate voltage $V_{\rm{G}}$. The capacitively coupled sample is represented as an Ohmic resistor $R$. The CPS is addressed using two-tone spectroscopy ($\omega_{\rm{r}}, \omega_{\rm{s}}$) in a reflective measurement scheme via a coupling capacitance $C_{\rm{in}}$. (c) The charge dispersion of the two lowest eigenstates is shown as a function of the gate charge $n_{\rm{G}}$. The Josephson energy $E_{\rm{J}}$ lifts their degeneracy at $n_{\rm G}=0$ and induces a level splitting $\Omega_0$ between $\ket{\uparrow}$ and $\ket{\downarrow}$. This realizes a two level quantum system.}
\label{fig1}
\end{figure*}

Dissipative transport characteristics in a sample can be probed using a CQ with the help of a coupling capacitor $C_{\rm{C}}$ (Fig.\,\ref{fig1}(a)). The capacitance serves as a mediator, effectively converting voltage fluctuations $\delta V$ in the sample into charge noise on the qubit. For sufficiently large $C_{\rm{C}}$ values, voltage fluctuations can therefore induce decoherence of a prepared CQ state. Decoherence of the CQ (and generally any two-level quantum system) resulting from that interaction can be quantified my measuring its decoherence time $T_{2}$. The $T_{\rm 2}$ time can therefore be used as a measurable quantity to probe dissipative processes in the immediate qubit environment, e.g. by using microwaves in a circuit quantum-electrodynamics (cQED) setup \cite{wallraff_2004}. Importantly, a cQED realization with GHz bandwidth also lends itself to time-resolved studies using a pulsed measurement scheme, which could facilitate the investigation of dissipative transport dynamics under non-equilibrium conditions with, in principal, picosecond temporal resolution \cite{zopes_2017}.

Realizing this noise spectroscopy concept in a local probe scenario requires a geometric coupling capacitance, the lateral extent of which is small enough to investigate dissipative charge carrier dynamics spatially resolved. Unlike in scanning NV microscopy, where the point-like NV quantum sensor brought in the vicinity of a sample facilitates a spatial resolution on the order of 10 nm \cite{degen_2008, maze_2008}, superconducting charge qubits are typically as large as tens of micrometers. Achieving spatial resolution therefore demands for a suitable coupling concept to their charge degree of freedom. Following other scanning probe techniques, we propose to use thin tips made from superconducting wire (diameter $\leq5\;\mu$m), to realize a local geometric coupling capacitance on the order of femtofarad to a sample surface underneath (see Fig.\,1(a)) \cite{ast_2016}. In this scheme we propose, the wire itself is attached and galvanically coupled to one of the CQ capacitor pads. This coupling idea of using the geometric capacitance of a thin tip is key to our proposal and overcomes the existing spatial resolution limit achieved with superconducting qubits owing to their macroscopic size for the application as local quantum sensors \cite{shanks_2013}.

\textbf{Decoherence in a charge qubit.} In the most simple implementation, the CQ corresponds to a charge island formed between a Josephson junction, characterized by the coupling energy $E_{\rm{J}}$ and its capacitance $C_{\rm{J}}$, and a gate capacitance $C_{\rm{G}}$, which allows to adjust the island charge $n_{\rm{G}}=C_{\rm{G}}V_{\rm{G}}/2e$ in units of Cooper pairs through applying a gate voltage $V_{\rm{G}}$ ($e$ - elementary charge) (Fig.\,1(b)). In the limit of the charging energy $E_{\rm{C}}=e^2/2C_{\Sigma}$ exceeding the Josephson coupling energy $E_{\rm{C}}\gg E_{\rm{J}}$, the system can be reduced to the two lowest charge states of the island $\ket{\uparrow}$ and $\ket{\downarrow}$, respectively. $E_{\rm{C}}$ is determined by the total capacitance to ground as $C_{\Sigma}=C_{\rm{J}}+C_{\rm{G}}$ -- and possible other contributions, such as the coupling capacitance $C_{\rm{C}}$. In this case, the effective Hamiltonian can be rewritten in the form of a fictitious spin-1/2 particle, $H=-E_{\rm{el}}/2 \sigma_{\rm{z}}-E_{\rm{J}}/2 \sigma_{\rm{x}}$, under the influence of the pseudo-magnetic fields $B_{\rm{z}}=E_{\rm{el}}$ and $B_{\rm{x}}=E_{\rm{J}}$ with the electrostatic energy $E_{\rm{el}}=4E_{\rm{C}} (1-2n_{\rm{G}})$ \cite{blais_2004}. 

The charge dispersion of $H$ is shown in Fig.\,1(c) and illustrates the role of $E_{\rm{J}}$ as a symmetry breaking term that lifts the degeneracy of $\ket{\uparrow}$ and $\ket{\downarrow}$ states at a gate charge of $n_{\rm{G}}=0.5$. This results in a gapped excitation spectrum with the level splitting $\Omega_0=\sqrt{E_{\rm{el}}^2+E_{\rm{J}}^2}$. For a given gate charge, the CQ therefore corresponds to a two level quantum system, the coherent superposition of which can be mapped onto a Bloch sphere (cf. Fig.\,1(a)).

Interaction with a dissipative environment can lead to decoherence and relaxation of a prepared CQ state, the strength of which is determined by the CQ properties and the coupling strength between the CQ and the environment. Voltage noise $\delta V(t)$ present in the environment is converted to gate charge noise $\delta n_{\rm{G}}(t)$ through the coupling capacitance $C_{\rm{C}}$. In the limit of $E_{\rm{C}}\gg E_{\rm{J}}$ this process will, in first order, only result in longitudinal charge fluctuations ($\delta n_{\rm{G}}(t)||\sigma_{\rm{z}}$), which can be rationalized in terms of a coupling Hamiltonian $H_{\rm{C}}(t)=4E_{\rm{C}} \sigma_{\rm{z}} \delta n_{\rm{G}}(t)$ \cite{schoelkopf_2003}. Under the influence of this noise term, the prepared quantum superposition of a CQ state will experience dephasing described by $\Delta\phi=\int_{-\infty}^{\infty}dt H_{\rm{C}}(t)$ \cite{weiss_1999}.

\begin{figure}[h!]
\centering
\includegraphics[height=8.6cm, width=8.6cm]{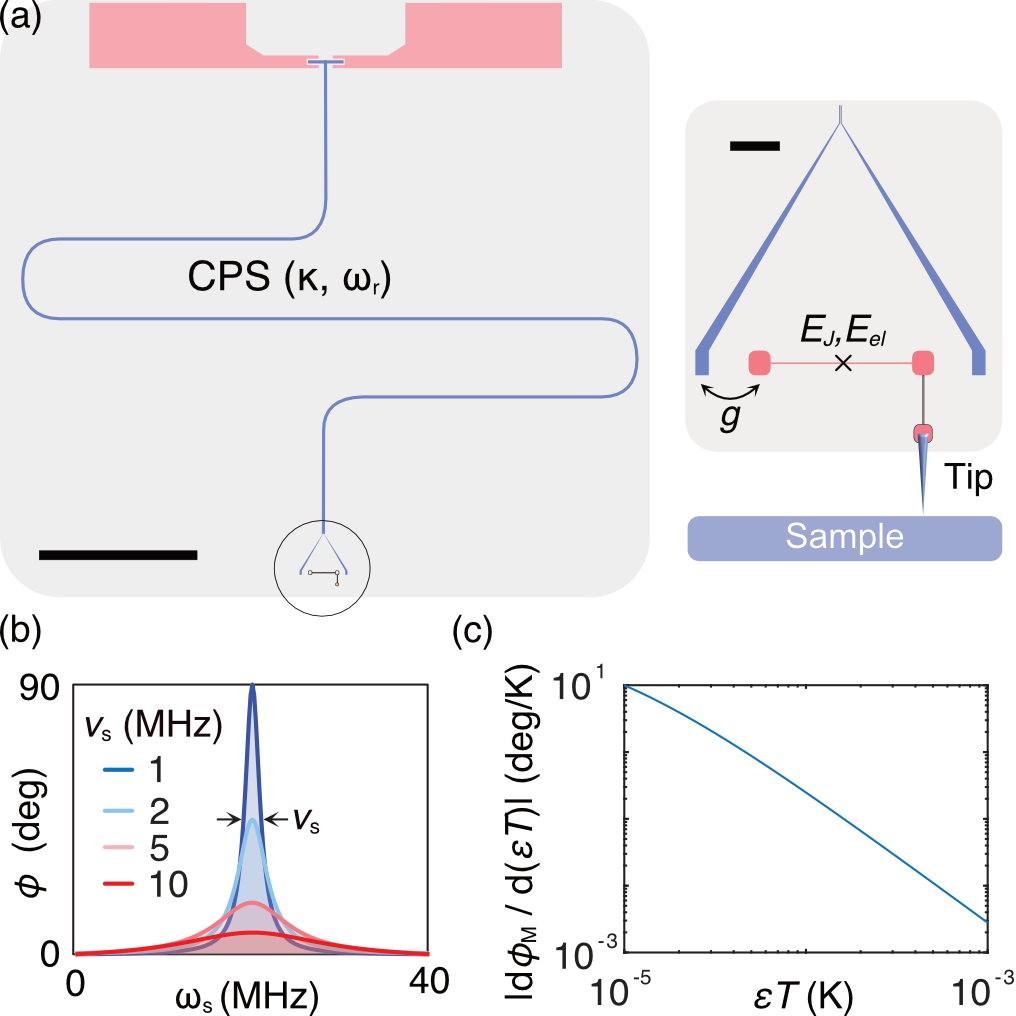}
\caption{\textbf{Microwave read-out of CQ decoherence.} (a) Realistic design drawing to scale of a resonator charge qubit implementation for SCQM. The red patches are the launchers for coupling to the circuitry, the coplanar strip-line (CPS) is shown in blue is characterized by its photon decay rate $\kappa$ and resonance frequency $\omega_{\rm{r}}$ (scale bar 800\,$\mu$m). Inset: The CPS couples to the qubit, characterized by its Josephson energy $E_{\rm{J}}$ and charging energy $E_{\rm{el}}$, through coupling capacitances (red pads) with strength $g$. The tip for realizing local probe capability is galvanically coupled to one of the capacitance pads (scale bar 50\,$\mu$m). (b) Calculated homodyne phase shift $\phi(\omega{\rm s})$ of the cavity tone as a function of the applied spectroscopy tone $\omega_{\rm{s}}$ for different indicated linewidths $\nu_{\rm{s}}$. (c) Calculated change of the phase shift peak maximum $\phi_{\rm{M}}$ in (b) with respect to changes in the sample's temperature $T$ times normalized resistance $\epsilon$.}
\label{fig2}
\end{figure}

On this basis, we can quantify the dephasing of a CQ state induced by Johnson noise in the environment. The magnitude of the resulting charge noise spectral density on the qubit, $S_{\rm{n}}=\zeta(\omega) S_{\rm{V}}$, is determined by the transfer function $\zeta=(\eta C_{\rm{C}}/2e)^2$, which for simplicity we assume to be frequency-independent. The scaling factor $\eta=C_{\rm{C}}/C_{\rm{\Sigma}}$ renormalizes $C_{\rm{C}}$ to an effective value with respect to the total capacitance to ground $C_{\rm{\Sigma}}$. Evaluating the dephasing $\Delta\phi$ in terms of the phase-phase correlation function in the low frequency limit \cite{weiss_1999}, one can write the charge noise induced dephasing time as,
\begin{equation}
    1/T_{\rm{2M}}=8\pi^2 (k_{\rm{B}} T/\hbar) \epsilon\eta^2
    \label{Eq1}
\end{equation} 
 with $\epsilon=R/R_{\rm{Q}}$ as the normalized resistance and $R_{\rm{Q}}$ as Klitzing's constant \cite{SI}. Measuring the CQ dephasing characteristics therefore allows to directly determine the resistance $\epsilon$ and temperature $T$ of a sample, which is capacitively coupled to the CQ. It bears noting that the response of the dephasing constant, $2\pi\nu_{\rm{M}}=1/T_{\rm{2M}}$, with respect to these parameters, $\partial\nu_{\rm{M}}/\partial(\epsilon T)=4\pi(k_{\rm{B}}/\hbar) \eta^2$, is constant, rendering the CQ an ideal quantum sensor with linear output characteristics. 

\begin{figure*}[h!]
\centering
\includegraphics[height=5.2cm, width=17.2cm]{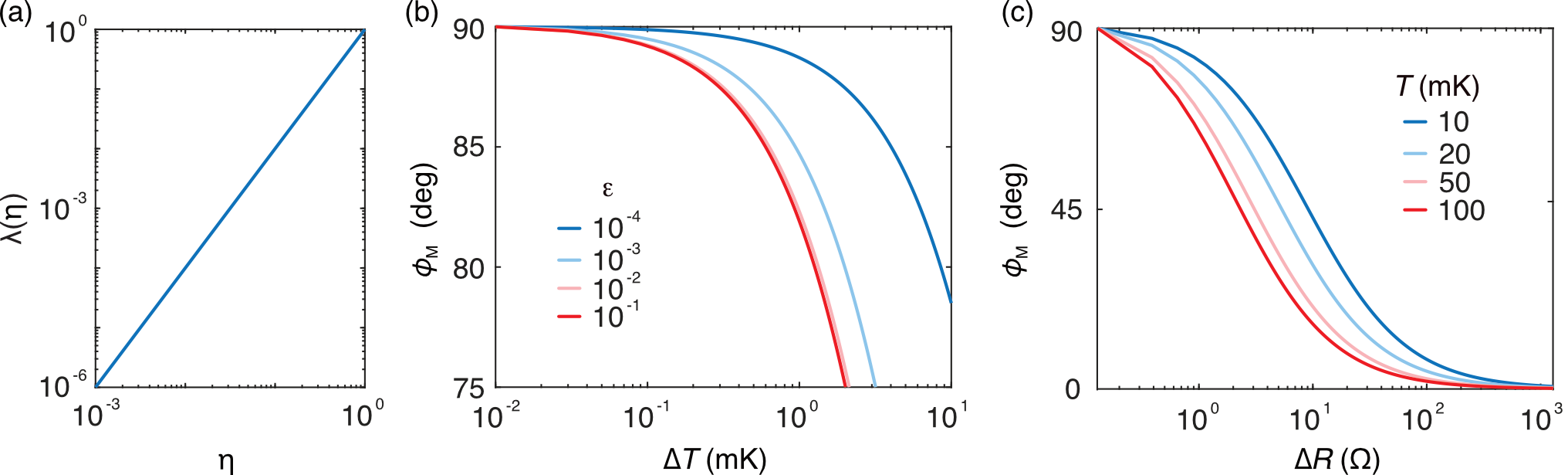}
\caption{\textbf{Response the SCQM quantum sensor.} (a) Normalized calculated response of the CQ quantum sensor, $\lambda=\partial \phi_{\rm M}/\partial(\epsilon T)$, to changes in the resistance, $\epsilon$, and temperature, $T$, of a sample as a function of the renormalized coupling capacitance $\eta$. Shown is the calculated change of the phase dip maximum, normalized to its maximum value at $\eta=1$. (b) Calculated homodyne phase shift, $\phi_{\rm M}$, as a function of a temperature change, $\Delta T$, in a capacitively coupled sample using realistic setup parameters ($\eta=0.1, \nu_{\rm{\phi}}=320\,$kHz) for different indicated values of the normalized sample resistance $\epsilon$. (c) Calculated homodyne phase shift, $\phi_{\rm M}$, as a function of resistance changes $\Delta R$ in a capacitively coupled sample using realistic setup parameters ($\eta=0.1, \nu_{\rm{\phi}}=320\,$MHz) and different indicated values of the temperature $T$.}
\label{fig3}
\end{figure*}

\textbf{Microwave control and read-out of a CQ.} The qubit state and its dephasing characteristics can be interrogated using microwave photons in a cQED architecture \cite{wallraff_2004, blais_2004}. In the example shown in Fig.\,2(a), the qubit is capacitively coupled to a superconducting coplanar stripline resonator (CPS) of bare frequency $\omega_{\rm{r}}$, realizing a reflective read-out scheme. In the limit of strong coupling $g\gg\kappa,\gamma$, in which the resonator-qubit coupling $g$ exceeds both the inverse cavity and the qubit lifetime, $\kappa$ and $\gamma$, respectively the CQ states ($\ket{\uparrow}$ and $\ket{\downarrow}$) are entangled with the resonator photon number states $\ket{n}$. At large detuning values, $\Delta=\omega_{\rm{r}}-\Omega_0$, between resonator frequency and CQ level splitting, the entanglement yields a dressed resonator frequency, $\omega_{\rm{r}}=\omega_{\rm{r}}\pm g^2/\Delta$, which depends on the CQ state (+, $\ket{\uparrow}$ and -, $\ket{\downarrow}$). This so-called dispersive limit allows to determine the CQ state by measuring the phase shift $\phi=\pm\tan^{-1}(2g^2)/(\kappa\Delta)$ of the reflected microwave photons when driving the cavity at its bare frequency $\omega_{\rm{r}}$ \cite{blais_2004}. 

Applying an additional spectroscopy tone, $\omega_{\rm{s}}$, to the CPS allows to prepare and probe an arbitrary coherent superposition of the $\ket{\uparrow}$ and $\ket{\downarrow}$ state. Importantly, such two-tone spectroscopy yields a strong (no) phase shift for the reflected microwave signal in a homodyne detection scheme, when $\omega_{\rm{s}}$ is on (off) resonance with the qubit transition. This is illustrated in Fig.\,2(b), which shows a calculated spectroscopy tone sweep across the $\ket{\uparrow}$-$\ket{\downarrow}$ transition. Hence, such two-tone spectroscopy is suited to both control and read-out the state of a CQ in our quantum sensing scheme.

Most commonly, pulse sequence experiments, such as Ramsey fringe or spin-echo, are used to measure the $T_2$ time with high accuracy \cite{degen_2017}. Determining the qubit dephasing through observing the quantum state evolution at different pulse delays in these experiments typically requires long measurement times on the order of minutes to hours. In the context of SCQM, this approach appears less suited from a practical perspective, as mapping out $T_2$ on a $256\times256$ point grid on a sample surface would result in measurement times on the order of a day or more. 

\textbf{Fast read-out of CQ decoherence.} We here propose an alternative approach for determining the dephasing characteristics that lends itself to fast measurement schemes and is key to this SCQM proposal. It is based on a line shape analysis of the spectroscopy tone sweep shown in Fig.\,2(b), facilitating the direct determination of the CQ's $T_2$ time. In the low power limit of $n_{\rm{S}}\rightarrow0$, in which only few spectroscopy photons, $n_{\rm{S}}$, are occupying the resonator at $\omega_{\rm{s}}$, the line shape of the phase peak can be approximated by a Lorentzian function, $\phi(\omega_{\rm{s}},\nu_{\rm{s}})$, the half-width half-maximum of which, $\nu_{\rm{s}}$, is directly linked to $T_2$ through $2\pi\nu_{\rm{s}}=1/T_2$ \cite{schuster_2005}. Crucially, owing to its Lorentzian nature, the maximum (minimum) of the phase peak (dip) is directly proportional to the dephasing time $\phi_{\rm{M}}=1/(\pi\nu_{\rm{s}})=1/(2\pi^2) T_2$. In the low power limit, $n_{\rm{S}}\rightarrow0$, it is therefore possible to determine the CQ's $T_2$ time through measuring $\phi_{\rm{M}}$ at a fixed applied spectroscopy tone frequency $\omega_{\rm{s}}$, set to be on resonance with the CQ transition $\Omega_{\rm 0}$. 

We want to point out that the reciprocal dependence of $\phi_{\rm{M}}\propto(\epsilon T)^{-1}$ on the resistance and temperature, renders this detection scheme highly responsive to even smallest changes in these quantities, as illustrated in Fig.\,2(c). Finally, this detection scheme, utilizing two tone spectroscopy at fixed frequencies, realizes a fast read out of the qubit decoherence characteristics, without the requirement of frequency or time delay sweeps. Hence, it facilitates the imaging of local dissipative transport properties on a sample surface in a realistic experimental time frame, which we will discuss in more detail below.

\textbf{Coupling capacitance and spatial resolution.} The ability of SCQM to measure dissipative transport dynamics in a capacitively coupled sample depends on the intrinsic CQ properties, the SCQM device architecture and the specific coupling geometry. Crucially, the detection of dissipative transport induced qubit decoherence requires that its rate, $\nu_{\rm{M}}$, exceeds the intrinsic qubit dephasing rate, $\nu_{\phi}$, in order to yield a measurable signal in the total dephasing signal, $1/T_2 =2\pi\nu_2=2\pi (\nu_{\rm{M}}+\nu_{\phi})=1/(T_{\rm{2M}}+1/T_{\phi})\approx1/T_{\rm{2M}}$ for $\nu_{\rm{M}}\gg \nu_{\phi}$. The typical intrinsic dephasing time of superconducting CQ, mostly determined by $1/f$-charge, noise is about $\nu_{\phi}\approx320\, $kHz \cite{schuster_2005}. This values determines the lower bound for the minimum required external dephasing rate $\nu_{\rm{M}}$ and thus, defines realistic boundary conditions for the experimental design.

The response $\lambda$ of the CQ to changes in the sample's temperature and resistivity, $\lambda=\partial\nu_{\rm{M}}/\partial(\epsilon T)\propto\eta^2$, can be significantly enhanced by maximizing the coupling capacitance $C_{\rm{C}}$ and, at the same time, by minimizing the total capacitance to ground  $C_{\Sigma}$, such that $\eta\rightarrow1$. Fig.\,3(a) illustrates that enhancing the renormalized coupling capacitance $\eta$ by one order of magnitude already increases $\lambda$ by two orders of magnitude. We have performed electrostatic simulations of the resonator-CQ-tip-sample geometry (Fig.\,2(a)) with the goal of minimizing $C_{\Sigma}$ and obtaining a realistic estimate for $\eta$ \cite{SI}. Our simulations show that using an optimized CPS design can reduce the total capacitance $C_{\Sigma}$ to ground to values as small as $C_{\Sigma}<3\,$fF. At the same time, large CQ resonator coupling values, $g\gg100\,$MHz, required for the dispersive readout scheme, $g\gg \kappa, \gamma$ can be maintained \cite{SI}. We note that our simulation results on the capacitance values are comparable to previously reported experimental values for Cooper pair boxes \cite{schuster_2005}. Overall, key CPS design aspect is to minimize the surface area of the capacitor pads that couple the Josephson junction to the resonator (Fig.\,2(a) inset).

Concerning the realization of sufficiently large $C_{\rm{C}}$ values, one has to find the right balance between, on the one hand maximizing the geometric capacitance for obtaining a high response and, on the other hand maintaining a high spatial resolution. Using a thin conical wire of base diameter $5\,\mu$m as a tip attached to one of the capacitor pads (see Fig.\,2(a) inset), our analytical calculations show that renormalized coupling capacitance values of $\eta\approx0.1$ and an effective spatial resolution of $\Delta x\geq100\,$nm can be achieved \cite{SI}. In this case, the spatial resolution is limited by the geometrically distributed stray capacitance of the tip wire. Using a thin superconducting nano-wire with a diameter of $50\,$nm as an alternative still allows to reach values of $\eta>0.01$ but with a much enhanced spatial resolution of $\Delta x\leq50\,$nm \cite{SI}. Ultimately, one can choose an appropriate coupling tip based on the requirements of an experiment for the spatial resolution and the sensor response.

\begin{table*}
\centering
\caption{\textbf{Scanning probe techniques for studying dissipative transport properties.} Comparison between quantum sensing and related measurement techniques, which probe dissipative charge carrier transport with spatial $\Delta x$ and temporal $\Delta\tau$ resolution, respectively. Relevant properties and measurement quantities are listed.}
\begin{tabular}{| l | l | l | c | c |}
\hline
Technique	 & Operating range (K) & Quantities (Resolution) & $\Delta x$ & $\Delta\tau$ \\ 
\hline
Scanning NV \cite{ariyaratne_2018} &	4 - 300 & $\sigma$  ($1\cdot10^{-4}\,\Omega^{-1}\cdot$cm$^{-1}$) & $40\,$nm & ps \cite{zopes_2017} \\ 
 & & $T$ ($\leq5\,$mK) & & \\
\hline
tSOT	 \cite{halbertal_2016}& 0.3 - 10	 & $T$ ($\leq1\,\mu$K) & $>50\,$nm &	- \\
\hline
MIM	\cite{cui_2016} & 2- 300 & $\rho$	& $100\,$nm & - \\
\hline
SCQM & $\leq\,0.1$ & $\rho$ ($\leq1\cdot10^{4}\,\Omega\cdot$cm) & $\leq50\,$nm & ps \cite{zopes_2017} \\
 & & $T$ ($\leq0.1\,$mK) & & \\
\hline
\end{tabular}
\label{tbl1}
\end{table*}

\textbf{Response and sensitivity of SCQM.} Based on these considerations it is possible to calculate the response of the CQ to changes in temperature, $\Delta T$ and resistance, $\Delta R$, of a capacitively coupled sample, that is the change of the detected phase maximum with respect to these quantities (cf. Fig.\,2(b)) \cite{SI}. The calculated response $\phi(\Delta T)$ of the CQ to a temperature change is displayed Fig.\,3(b) and illustrates the high sensitivity of the CQ to even smallest changes in temperature in this experimental concept. Using realistic setup parameters (see caption), our calculations reveal that a temperature change of $\Delta T=1\,$mK, such as that induced by dissipative charge carrier dynamics in a sample, already induces a $10\,\%$ change in the measured homodyne phase shift. The calculated response to a resistance change $\phi(\Delta R)$ in Fig.\,4(b) displays similar sensitive characteristics -- a change of $\Delta R=10\,\Omega$ reduces the phase maximum by $\approx50\,\%$.

Ultimately, the signal-to-noise ratio (SNR) and the attainable sensitivity of SCQM to temperature and resistance changes are determined by the noise level of the amplification line for the microwave signal. Its noise is commonly dominated by the number of thermal photons $n_{\rm{D}}=k_{\rm{B}} T_{\rm{D}}/(\hbar\omega_{\rm{r}}$ generated in the high-electron-mobility-amplifier (HEMT) used to amplify the microwave signal at cryogenic temperatures ($T_{\rm{D}}=4\,$K). A conservative estimate of that number for our concept yields $n_{\rm{D}}\approx100$. In a homodyne detection scheme at $g^2/(\kappa\Delta)\gg1$, we can then define the $SNR=\sqrt{m}(n_{\rm{P}}/n_{\rm{D}})$, where $m$ corresponds to the number of measurements and $n_{\rm{P}}=n\kappa\tau/2$ to the number of collected photons during a finite integration time $\tau$. 

If we assume a cavity enhanced CQ lifetime of  $1/\gamma=(\Delta/g)^2 \kappa^{-1}\approx100\,\mu$s and operate our read-out in the low power limit $n\leq10$, we can define a theoretical upper bound to the attainable $SNR=5\cdot\sqrt{m}$, for $\tau=1/\gamma$. Considering an experimentally determined CQ lifetime of $1/\gamma\approx2\,\mu$s for superconducting charge qubits \cite{schuster_2005}, we obtain a realistic SNR estimate of $SNR=10^{-1}\cdot\sqrt{m}$. Hence, with large amount of averaging ($m>10^6$) a $SNR>100$ can be realized, which should facilitate a temperature and resistivity resolution of $\Delta T\leq0.1\,$mK and $\Delta R\leq0.1\,\Omega$. Assuming a tip diameter of $50\,$nm, a change in resistivity of $\Delta\rho\leq1\cdot10^{4}\,\Omega\cdot$cm could be resolved. 

Fast scanning operation of SCQM, on the other hand ($256\times256$ grid, total measurement time $t<30\,$min), can be realized with moderate averaging ($m=10^4$). This results in $SNR=10$, which still allows to detect temperature and resistivity changes as small as $\Delta T\approx1\,$mK and  $\Delta\rho\leq5\cdot10^{4}\,\Omega\cdot$cm. If a higher sensitivity of the CQ sensor is required, using Josephson parametric amplifiers instead of HEMTs could enhance the SNR by up to two orders of magnitude \cite{macklin_2015}.

\textbf{Scientific use cases for SCQM.} To discuss the perspectives of SCQM in the context of quantum materials, it is instructive to review the current state of related measurement techniques. Table\,\ref{tbl1} contrasts the capabilities of some of the existing microscopy techniques to probe dissipative transport with the estimated performance of SCQM. Among those techniques, scanning NV microscopy represents the most versatile tool with demonstrated capabilities of tracking a sample's electrical conductance and temperature with nanometer spatial and, in principal, picosecond temporal resolution across a large temperature range down to $4\,$K \cite{ariyaratne_2018}. Thermal imaging using a SQUID on tip (tSOT) operating at $T\geq300\,$mK, offers DC thermal imaging capabilities with an unprecedented temperature resolution of $\Delta T\leq1\,\mu$K and a spatial resolution of $\approx100\,$nm \cite{halbertal_2016}. Scanning microwave impedance microscopy is specialized on resistivity imaging with approx. 100\,nm spatial resolution at  $T>2\,$K \cite{cui_2016}. In comparison, SCQM offers resistivity and temperature imaging with an estimated spatial resolution of better than $50\,$nm and a theoretical temporal resolution on the order of picoseconds \cite{zopes_2017}. It has therefore potential capabilities similar to those of scanning NV microscopy in terms of resistivity and temperature resolution, but operates in a lower temperature window much below $1\,$K.

Put into context of quantum materials, SQCM will therefore be especially suited for a number of different applications. It could provide microscopic insight on charge carrier interaction driven quantum phase transitions in correlated phases of matter, occurring at temperatures below $1\,$K. Examples would be the superconductor-insulator-transition found in magic angle twisted bilayer Graphene \cite{cao_2018_1, cao_2018_2} and in monolayer $\rm{WTe}_2$ \cite{wu_2018, fatemi_2018}. Through its potential to distinguish between different transport regimes, SCQM should also be of value to detect and study the transport characteristics of topologically protected boundary states in novel higher order topological insulator platforms \cite{benalcazar_2017, xu_2019}, and to shed light on hydrodynamic transport and the underlying mechanisms, too \cite{bandurin_2016, crossno_2016, moll_2016}. In this context, dynamic decoupling pulse sequences could be used to determine the full frequency-dependent spectrum of the noise spectral function $S_{\rm V}(\omega, T, R)$ \cite{degen_2017}. Owing to the potential high temporal resolution \cite{zopes_2017}, SQCM could also help to shed light on charge carrier dynamics in a variety of mesoscopic and nanoscale devices, such as the non-equilibrium quasiparticles in superconducting films, which are known to deteriorate the performance of superconducting qubits \cite{riste_2013, serniak_2018}. In a broader context, the noise spectroscopy based on qubit decoherence described above will also facilitate the investigation of dissipative transport characteristics of 2D quantum materials in non-local experiments using pure on-chip realizations (cf. Ref.\,\cite{kolkowitz_2015}).

\textbf{Technical aspects.} As far as the technical feasibility of SCQM is concerned, it is designed around established fabrication and measurement techniques for qubits and off-the-shelve technology for the scanning module and cryostat environment. The design and fabrication of superconducting qubits has experienced a tremendous development over the past decades \cite{devoret_2013}, promising an optimized performance beyond reported Cooper pair box results of $1/\gamma\approx2\,\mu$s \cite{schuster_2005}, for instance using a larger $E_{\rm J}/E_{\rm C}$ ratio \cite{koch_2008}. We note that gate charge drift, which changes the level splitting $\Omega_0$ over the course of minutes and is a known weak spot of Cooper pair boxes, can be compensated for by feedback mechanisms and, thus, should not interfere with the SCQM operation. Moreover, low loss CPS resonators with internal quality factors $Q_{\rm{i}}>10^5$ in the limit $n_{\rm{s}}\rightarrow0$ can be reliably fabricated from different materials nowadays \cite{sage_2011}, helping to satisfy the condition $g>\kappa$ in the more complex environment of a SCQM setup ($\kappa=\omega_{\rm{r}}/Q_{\rm{i}}$). Potential Purcell losses into the DC lines connecting the sample owing to the coupling capacitance $C_{\rm C}$ can be mitigated by appropriate on- and off-chip filtering \cite{bronn_2015}. 

Scanning operation can be readily implemented using commercially available nano-positioners, suitable to operate at lowest temperatures. Using qPlus sensor technology may, in addition, enhance the scanning operation of SCQM \cite{giessibl_2019}. Previous studies have already demonstrated the feasibility of integrating scanning probe setups into dilution refrigerator cryostats while maintaining lowest electron temperatures \cite{shanks_2013, sueur_2008}, an aspect crucial to the operation of the CQ sensor. Yet, attention has to be paid to a proper thermalization and filtering of the DC lines needed for scanning operation \cite{assig_2013, scheller_2014, machida_2018}. Regarding the tip-to-qubit fabrication, fusing the tip to a capacitor pad could be performed by means of micro soldering or focused ion beam assisted deposition, an approach commonly practiced in the context of the most recent scanning NV tip technology \cite{maletinsky_2012}. Alternatively, one can also envision more sophisticated all-on-chip solutions in the style of microfabricated tips already existing for scanning tunneling microscopy applications \cite{allan_2019}.

\textbf{Conclusion.} We proposed SCQM as a new quantum sensing imaging modality to study dissipative transport dynamics of new electronic phases in low-dimensional quantum materials. Backed up by model calculations, we demonstrate design concepts for local probe realizations based on the geometric capacitance forming between a sample and a tip, which is coupled to the charge qubit. We propose a tangible scheme for fast microwave read out of the qubit decoherence using standard homodyne techniques, facilitating fast scanning operation and realistic measurement times for SCQM. Our analytical and numerical analyses reveal the potential capability of SCQM to resolve temperature and resistivity changes of a sample as small as $\Delta T\leq0.1\;$mK and $\Delta\rho\leq1\cdot10^{4}\;\Omega\cdot$cm, respectively. SQCM, therefore overcomes existing limitations of superconducting qubits for quantum sensing applications and will be especially suited to study the microscopic mechanism of interaction driven quantum phase transitions in low-dimensional correlated phases of matter, visualize the local transport characteristics of novel topological materials as well as to investigate dissipative charge carrier dynamics in quantum materials with high spatial and temporal resolution.

{\em Acknowledgements --} It is our pleasure to acknowledge inspiring and educating discussions with Andrew Houck, Ali Yazdani, Andras Gyenis, Alex Place and Yonglong Xie. This work has been primarily supported by the Alexander-von-Humboldt foundation through a Feodor-Lynen postdoctoral fellowship. Additional support was provided by the Office of Naval Research (ONR N00014-16-1-2391, ONR N00014-17-1-2784, ONR N00014-13-1-0661), ExxonMobil (EM09125.A1) and the Gordon and Betty Moore Foundation as part of the EPiQS initiative (GBMF4530). BJ acknowledges the hospitality of the Houck lab at Princeton University.

\end{document}